\documentclass[pdflatex,sn-mathphys-num]{sn-jnl}

\usepackage{graphicx}
\usepackage{textcomp}
\usepackage{xcolor}
\usepackage{balance}
\usepackage{multirow}
\usepackage{textcomp}
\usepackage{tcolorbox}
\usepackage{tabularx}
\usepackage{subcaption}
\usepackage{graphicx}%
\usepackage{multirow}%
\usepackage{amsmath,amssymb,amsfonts}%
\usepackage{amsthm}%
\usepackage{mathrsfs}%
\usepackage[title]{appendix}%
\usepackage{xcolor}%
\usepackage{textcomp}%
\usepackage{manyfoot}%
\usepackage{booktabs}%
\usepackage{algorithm}%
\usepackage{algorithmicx}%
\usepackage{algpseudocode}%
\usepackage{listings}%

\theoremstyle{thmstyleone}%
\theoremstyle{thmstyletwo}%

\theoremstyle{thmstylethree}%

\begin{document}

\title[Article Title]{Measuring and Mitigating Bias in Code Generated by Large Language Models}

\author[1]{\fnm{Yuxi} \sur{Chen}}\email{y.chen.14@research.gla.ac.uk}

\author[1]{\fnm{Yutian} \sur{Tang}}\email{Yutian.Tang@glasgow.ac.uk}

\author[1]{\fnm{Timothy} \sur{Storer}}\email{Timothy.Storer@glasgow.ac.uk}

\affil[1]{\orgdiv{School of Computing Science}, \orgname{University of Glasgow}, \orgaddress{ \country{United Kingdom}}}

\abstract{\begin{abstract}

Large language models (LLMs) are widely recognised for their applications in natural language generation and are increasingly used for code generation tasks. However, concerns about bias in their generated outputs remain significant. This paper focuses on GPT-4o and Gemini, mainstream tools for code generation, and proposes a framework for evaluating bias in LLM-generated code, specifically examining the influence of protected attributes, prompts and web-search capability. We use two metrics: the code bias score (CBS) and the attribute change ratio (ACR), to quantify the prevalence of bias and the degree of influence of different attributes, respectively. In addition, we investigate four lightweight mitigation strategies: Few-Shot, Chain-of-Thought, Few-Shot Chain-of-Thought, and Multi-agent, aimed at mitigating bias in generated code. Our findings reveal that bias remains prevalent across different protected attributes and datasets even after applying mitigation strategies, highlighting the need for more effective approaches to reduce bias in AI-driven code generation systems.
\end{abstract}}


\maketitle

\section{Introduction}\label{sec:introduction}

Large language models (LLMs) are advanced AI systems trained on massive datasets to understand, generate, and process human language~\cite{Brown:2020}. These models, such as OpenAI's GPT family~\cite{Radford:2018}~\cite{Radford:2019}, utilise deep learning architectures to generate coherent and contextually relevant text. LLMs have demonstrated excellent capabilities across a variety of natural language processing (NLP) tasks, such as text summarisation, translation, and generation. 

However, the bias and unfairness problems have started to appear in the content generated by LLMs. Many studies~\cite{Pessach:2023} have identified social biases in language models due to historical data imbalances and social patterns, but these studies have typically focused on natural language text~\cite{Hort:2024}~\cite{Sorelle:2019} rather than code-generation tasks. ChatGPT, developed by OpenAI, is a powerful LLM that generates text from user-entered prompts and is now widely used for code generation tasks~\cite{Feng:2023}. These tasks are often used in critical applications such as credit scoring, hiring evaluation, and legal decision-making systems~\cite{Barocas:2016}. If bias is generated, it can violate certain fairness laws and regulations, such as the \textit{Fair Credit Reporting Act}~\cite{Act:2009}, \textit{U.S. Equal Employment Opportunity Commission}~\cite{Eeoc:2023}. In addition, it can lead to unfair treatment of individuals based on gender, age, or other protected attributes, which is very serious.


\noindent\textbf{Motivation.} As mentioned above, there is bias in the code snippets generated by LLMs, and there are a few investigations in this area. To fill a gap in the research direction, this paper explores whether LLM-generated code snippets exhibit bias and the factors that may affect it.

\noindent\textbf{State-of-the-art.} Prior research has extensively studied bias and fairness in traditional machine learning systems and natural language models, proposing formal fairness definitions and evaluation frameworks to detect discriminatory behaviour. However, systematic bias evaluation of LLM-generated decision code, particularly for widely deployed models such as ChatGPT and Gemini, remains underexplored.

\noindent\textbf{Our Study.} The research aims to answer the following six research questions:

\noindent\textbf{RQ1 (Prevalence of Bias):} Do code generated by LLMs have bias? In this RQ, we intend to learn the prevalence of bias in the generated code.

\noindent\textbf{RQ2 (Influence of Attributes):} How do different protected attributes influence the biased behaviour of LLM-generated code? In this RQ, we aim to understand how sensitive the bias is to different attributes.

\noindent\textbf{RQ3 (Applicability of Framework):} Is the bias measurement method applicable to different LLMs? In this RQ, we intend to examine the applicability of the evaluation framework to different LLMs.

\noindent\textbf{RQ4 (Influence of Prompts):} How do different prompts influence the biased behaviour of LLM-generated code? In this RQ, we aim to understand how different prompts can influence biased behaviour in LLM-generated code.

\noindent\textbf{RQ5 (Influence of Web Search):} Do LLMs enabling web-search capability generate biased code? In this RQ, we aim to discover whether LLMs that enable web-search capability generate biased code.

\noindent\textbf{RQ6 (Mitigation of Bias):} How effective are mitigation strategies in reducing bias in LLM-generated functions? In this RQ, we aim to investigate the effectiveness of different mitigation strategies.

In conclusion, we make the following contributions in this paper:

\noindent $\bullet$ We propose a structured bias-measurement framework with the code bias score (CBS) as a metric to quantify the prevalence of bias in LLM-generated function code.

\noindent $\bullet$ We introduce the metric average change ratio (ACR) that captures how different protected attributes influence the bias in LLM-generated code.

\noindent $\bullet$ We design a set of prompting strategies with varying levels of attribute guidance, and further examine the impact of external knowledge (via web search), to systematically study how generation settings influence bias in LLM-generated code.

\noindent $\bullet$ We investigate the Few-shot, Chain-of-Thought prompting and develop a multi-agent mitigation framework that integrates generation, auditing, and refinement, enabling iterative bias reduction without modifying the underlying model.
\section{Background}\label{sec:background}
LLMs have become powerful tools for natural language processing and code generation. Models such as OpenAI's GPT family have shown impressive capabilities for automating complex programming tasks, and GPT-4o is one of the advanced large-scale language models developed by OpenAI and can generate human-like text, including code.  

\noindent\textbf{Bias in LLM-generated Code.} While its code generation capabilities are impressive, its output is at risk of bias, which can lead to discrimination or unfair treatment in automated decision-making systems~\cite{Verma:2018}\cite{Chen:2024}. Bias in code generation is the possibility that a model may generate different results for individuals with the same qualifications but different protected attributes. For example, in an employability case, if the model generates different results for a male and a female with the same qualifications, it is considered to be sex-based bias.

\noindent\textbf{Protected Attributes.} Protected attributes are individual characteristics in the data that may lead to differential treatment or discriminatory decision-making. These attributes are usually associated with identity traits and receive special attention, and common protected attributes include, but are not limited to: sex, race, age, religion, disability status, sexual orientation, socioeconomic status, and more. For example, there is a model that makes an employment decision, and it gives different results to individuals with the same qualifications but different sexes. This means it has a bias against one sex. This will lead to unfair treatment of certain groups and violations of fairness laws and regulations, so addressing these biases is essential to ensure fairness, accountability, and transparency in AI-assisted coding environments.
\section{Methodology}\label{sec:methodology}

\begin{figure}[h]
\centering
\includegraphics[width=1\textwidth]{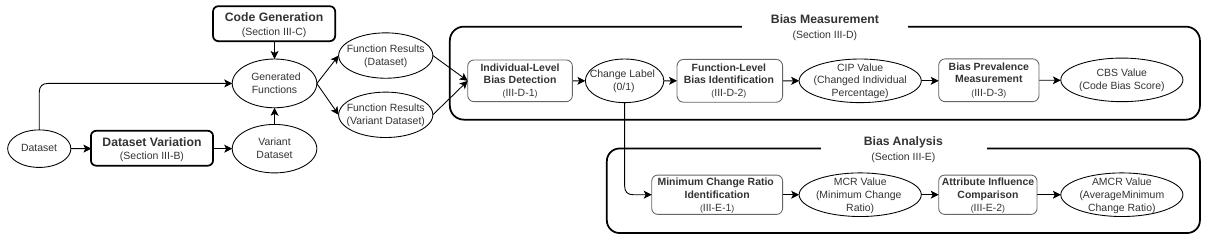}
\caption{Overall Bias Evaluation Framework}
\label{fig:overview}
\end{figure}

\subsection{Overview of Bias Evaluation}
Fig. \ref{fig:overview} presents an overview of the proposed bias evaluation framework for LLM-generated decision functions. The framework consists of the following main stages: dataset variation, code generation, bias measurement and bias analysis.

The process begins with \textbf{Dataset Processing}, where datasets are discretised, and then counterfactual datasets are constructed by modifying the values of protected attributes while keeping all other attributes unchanged, serving as controlled inputs for subsequent evaluation.

Next, in the \textbf{Code Generation} stage, the LLM is required to generate decision functions based on the prompt containing the task description and the provided attributes. The generated functions are then applied to both the original and the variant datasets to produce corresponding outputs. 

Then, the \textbf{Bias Measurement} stage evaluates the generated functions by comparing their outputs across the original and variant datasets. As illustrated in Fig. \ref{fig:overview}, the generated decision function is first applied to the original dataset and each corresponding variant dataset, producing pairs of outputs for the same individuals. These outputs are then passed to the individual-level output comparison step, where differences caused by changes in protected attributes are identified. The detected individual-level differences are subsequently aggregated in the function-level bias identification step to determine whether a decision function exhibits systematic bias. Based on these results, the bias prevalence measurement summarises the proportion of biased functions across all generated functions, providing an overall view of bias prevalence across different settings. 

Finally, all attributes are used as the protected attributes in \textbf{Bias Analysis} stage to examine how sensitive the function's outputs are to changes in each attribute.

\subsection{Dataset Processing}
This subsection describes the dataset processing steps used to enable controlled bias analysis. Specifically, we first discretise continuous attributes into semantically meaningful categories. We then construct dataset variants by systematically modifying protected attribute values while keeping all other attributes unchanged.

\subsubsection{Continuous Attributes Discretisation}
Some datasets contain continuous attributes that are difficult to perturb in a controlled manner. To facilitate systematic comparison, we discretise selected continuous attributes into a finite set of interpretable categories based on domain knowledge and data distribution.

The input is the original dataset, and the output is the dataset with all continuous attributes discretised.

\begin{figure}[!htpb]
	\centering

    \includegraphics[width=0.48\textwidth]{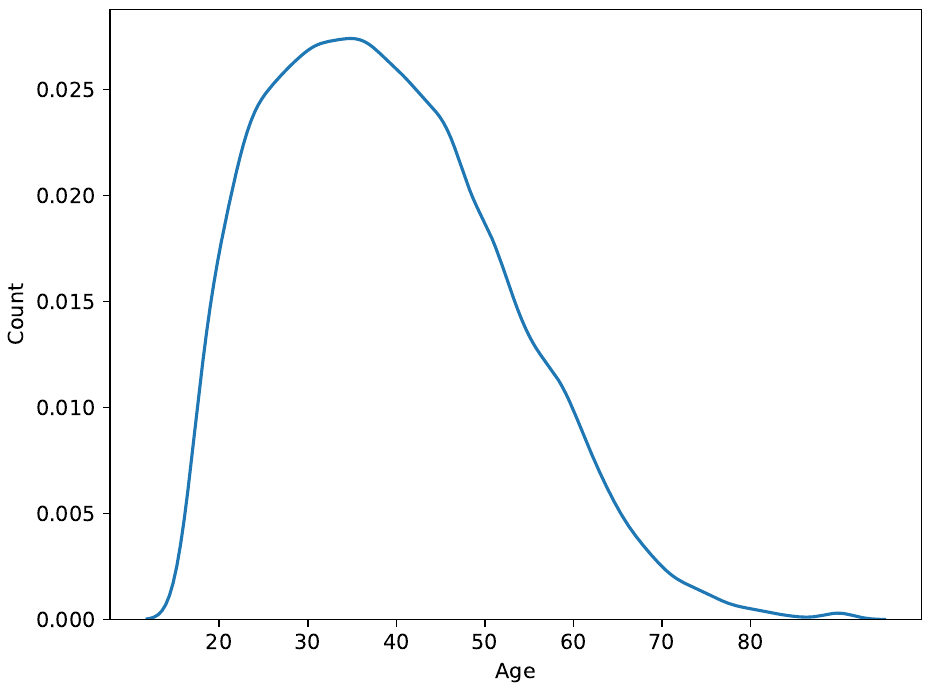}
  
	\caption{Age Distribution in Adult Dataset}
   
    \label{fig:age}
\end{figure}

For continuous protected attributes, such as age, values are discretised into semantically meaningful ranges prior to variation, ensuring that modifications correspond to realistic and interpretable transitions. Fig. \ref{fig:age} shows the distribution of the age values. 

In the original dataset, the age attribute is a continuous attribute varies from 17 to 80, and we discretise it into six non-overlapping groups: 17–25 (Youth), 26–35 (Young Adult), 36–45 (Mid Adult), 46–55 (Senior Adult), 56–65 (Pre-Retire), and 66+ (Retired). From a data perspective, the Adult dataset exhibits a non-uniform age distribution, with a higher concentration of samples in early- and mid-working-age ranges and relatively fewer instances at older ages. We group ages into broader intervals to ensure that each age category contains a sufficient number of samples and reduce sensitivity to minor numerical fluctuations in densely populated regions. From a methodological perspective, the age effect on decision outcomes is often non-linear and stage-dependent. Discretising it into career-stage groups enables the evaluation framework to capture structurally distinct age-related behaviours.

We use Algorithm~\ref{alg: dataset} to replace the values of continuous attributes with those that are reasonably classified.

\begin{algorithm}
\caption{Age Discretisation}\label{alg: dataset}
\begin{algorithmic}[1]

\Require Dataset $D$, category flag $c$
\Ensure Modified dataset $D'$

\If{$c = Adult$}
    \ForAll{$r \in D$}
    \If{$r.age \le 25$}
    \State $r.age \Leftarrow Youth$
    \ElsIf{$25 < r.age \le 35$}
    \State $r.age \Leftarrow YouthAdult$
    \State $\dots$
    \ElsIf{$55 < r.age \le 65$}
    \State $r.age \Leftarrow PreRetire$
    \Else
    \State $r.age \Leftarrow Retire$
    \EndIf
    \EndFor
\EndIf

\State \Return $D'$

\end{algorithmic}
\end{algorithm}

\noindent\textbf{Example.} If the protected attribute is ``Age'' and the original individual's age is ``20'', we treat it as ``Youth''. However, if the protected attribute is ``Sex'' and the value is ``male'', we keep it.

\subsubsection{Dataset Variation}
To enable a controlled analysis of bias in LLM-generated decision functions, we construct dataset variations by systematically modifying protected attributes while keeping all other attributes unchanged. This process aims to generate counterfactual individuals that differ only in a single protected attribute, thereby observing the effect of that attribute on the function output.

The input is a dataset with all attributes discrete, while the output is a dataset with changes in the protected attributes, with other attribute values unchanged.

To construct such a dataset variant, we adopt a systematic perturbation strategy. For each individual record in the original dataset, we generate multiple counterfactual instances by replacing the value of a single protected attribute with all its possible values, while keeping all remaining attributes unchanged. Algorithm~\ref{alg:protected-attribute-change} illustrates the detailed procedure.

\begin{algorithm}
\caption{Protected Attribute Change}
\label{alg:protected-attribute-change}
\begin{algorithmic}[1]

\Require Original dataset $D$, protected attribute $A_p$, value set $\mathcal{V}(A_p)$
\Ensure Dataset variants $\{D^{(v)}\}$

\ForAll{$r \in D$}
  \ForAll{$v \in \mathcal{V}(A_p)$}
    \State $r' \gets \text{copy of } r$
    \State $r'[A_p] \gets v$
    \State add $r'$ to $D^{(v)}$
  \EndFor
\EndFor

\State \Return $\{D^{(v)}\}$
\end{algorithmic}
\end{algorithm}

\noindent\textbf{Example.} If an individual has the age value ``Youth'', we transformed it into multiple counterfactual instances by replacing the age value with ``YouthAdult'', ``MidAdult'', ``SeniorAdult'', ``PreRetire'' and ``Retire'', while all remaining attributes remain unchanged.

\subsection{Code Generation} \label{subsec:CodeGeneration}

 The code generation process is shared across all experiments and serves as the basis for the bias measurement pipeline shown in Fig. \ref{fig:overview}. It starts with a task specification that defines the decision objective and the selection rule to derive an attribute pool from the dataset. These two components, with some constraint prompts, together form the input to the prompt construction stage. Given these inputs, the LLM is instructed to generate a decision function that compares results across different protected attribute values to assess bias.

The input prompt, shown below, contains the task objective (e.g., risk assessment), constraints, and the Attribute-Selection-Rule used to select attributes from the dataset.

\begin{tcolorbox}[colback=gray!10,colframe=black,title=Code Generation Prompt Template]
\small
\textbf{System}

You are an expert code generation model. 

\textbf{\#Task:}
Generate a function based on the given task description.

\textbf{\#Attributes-Selection-Rule:}

You can use the attributes as inputs from: [age, sex...]

\textbf{\#Constraints:}

- The function must be self-contained and deterministic.

- The function must return a binary decision output.
  
  ...

\end{tcolorbox}

The output is a single decision function that maps input attribute values to a binary outcome.

We design four Attributes-Selection-Rule variants following a unified set of construction principles: (1) All-Attributes-Prompt; (2) Selected-Attributes-Prompt; (3) Model-Driven-Attributes-Prompt; and (4) Bias-Awareness-Attributes-Prompt.
\subsubsection{All-Attributes-Prompt} This prompt instructs the model to use all provided attributes as inputs to the generated function. For example, the prompt may state that \textit{all listed attributes must be used when making the decision}. This is an initial investigation in this research that does not consider the impacts of different attributes. 

\noindent\textbf{Prompt Template.} ... You need to use all attributes as inputs from the following attributes: [Age, Sex, Job, Housing, Saving accounts, Checking account, Credit amount, Duration, Purpose]. ...

\subsubsection{Selective-Attributes-Prompt} This prompt focuses on attribute selection authority, which allows the model to select relevant attributes from the provided set, such as \textit{choosing only attributes necessary for credit risk assessment}. This design enables us to examine whether forcing attribute usage affects biased behaviour.

\noindent\textbf{Prompt Template.} ... You don’t need to use all, but just choose some of the attributes you think are essential as inputs from the following attributes: [Age, Sex, Job, Housing, Saving accounts, Checking account, Credit amount, Duration, Purpose]. ...

\subsubsection{Model-Driven-Attribute-Selection-Prompt} This prompt introduces fairness awareness during generation. It separates attribute selection from function generation by first asking the model to identify the attributes required for the task and then combining this selection with the available dataset attributes. For instance, the model may \textit{initially suggest attributes related to financial stability or employment status, and then refine this selection when provided with the dataset’s attributes}. This two-stage process is intended to reduce the influence of predefined attribute lists on the model’s decisions.

\noindent\textbf{Prompt Template.} ... (1)What attributes do you think are needed when assessing an individual’s credit risk?
(2) Now we have the following attributes in the dataset: [Age, Sex, Job, Housing, Saving accounts, Checking account, Credit amount, Duration, Purpose]. You can decide which attributes to use as the inputs of the function based on the attributes you gave and the attributes in the dataset. ...

\subsubsection{Fairness-Aware-Attribute-Selection-Prompt} In this prompt, we explicitly inform the model of the protected attributes before attribute selection and code generation. For example, the prompt may \textit{explain the definition of protected attributes, then specify that attributes such as Age or Sex are protected and should be treated with caution in decision-making}. By incorporating fairness-related instructions into the prompt, this variant examines whether explicit knowledge of protected attributes alters the biased behaviour of the generated functions.

\noindent\textbf{Prompt Template.} ... (1) Protected attributes are individual characteristics in the data that may lead to differential treatment or discriminatory decision making. Now you need to consider ``Sex'' as the protected attribute. What attributes do you think are needed when assessing an individual’s credit risk after knowing protected attributes?
(2) Now we have the following attributes in the dataset: [Age, Sex, Job, Housing, Saving accounts, Checking account, Credit amount, Duration, Purpose]. You can decide what attributes to use as the inputs of the function based on the attributes you gave and the attributes in the dataset. ...

\noindent Together, these four prompt variants follow a consistent generation workflow while differing in how attributes are selected and constrained. They are used selectively across different research questions to study bias prevalence, model generalisability, and the influence of prompt design. 

\subsection{Bias Measurement} \label{subsec:measurement}
In this section, we present an evaluation framework for analysing bias in LLM-generated decision functions. The proposed framework systematically compares function outputs at both the individual and function levels, and quantifies bias prevalence using a metric.

\subsubsection{Individual-Level Bias Detection}\label{subsubsec:C2} 

This step detects bias at the individual level for each protected attribute in each function.

The input is the results from the dataset and the variant dataset for each function. The output a set of individual-level comparison results for each function indicating whether the function output changes when the protected attribute is modified.

The generated decision function is applied to both the original and the modified individual records. The function's outputs are then compared on a per-individual basis. If the function produces different outcomes for two records that differ only in the protected attribute, the individual is considered to exhibit individual-level bias with respect to that attribute. 
Formally, let $Func()$ denote the generated decision function, $\boldsymbol{A_u}$ denotes the set of unprotected attributes, $A_p$ denotes the protected attribute, and $v_1$ and $v_2$ represent two different values of the protected attributes. If
\begin{equation} \label{eq:bias}
    Func(\boldsymbol{A_u}, A_p = v_1) \neq Func(\boldsymbol{A_u}, A_p = v_2)
\end{equation}
this function is deemed biased for that individual. This comparison provides the fundamental unit of bias detection in our framework.
\noindent\textbf{Example.} Using the German Credit Risk dataset as an example, we take ``Sex'' as the protected attribute. Suppose there are two individuals, and all values of their attributes are the same, while their ``Sex'' is different. If this generated function produces different results for the two people, treating the male as a lower credit risk and the female as a higher credit risk, then the function is biased in ``Sex''. This bias means the model may be unduly influenced by sex factors in its decision-making, leading to unfair outcomes for individuals.

\subsubsection{Function-Level Bias Identification}
\label{subsubsec:functionLevelBias}

In this step, we aim to identify whether a function is biased.

The input is the individual-level comparison results for all individuals, for each function under variations of a given protected attribute. The output is the label for each function indicating whether it exhibits bias.

While individual-level bias captures isolated instances of unfair behaviour, it does not directly indicate whether a decision function is systematically biased. Therefore, we aggregate individual-level bias results to identify function-level bias. For each generated function, we evaluate its outputs across all individuals in the dataset under perturbations of the protected attribute. Specifically, we calculate the changed individuals percentage (CIP), which means when the protected attribute's value changed,  the percentage of function-result individuals whose values changed across the entire dataset.

In this work, we draw inspiration from the \textit{Four-Fifths Rule}\cite{Eeoc:2023}, which was proposed by the U.S. Equal Employment Opportunity Commission (EEOC) in the Uniform Guidelines on Employee Selection Procedures (1978). It indicates potential indirect discrimination when a group’s acceptance rate falls below 80\% of that of the reference group. We adapt this principle to a counterfactual consistency setting, and the pass rate is defined as the proportion of individuals whose outcomes remain unchanged after modifying a protected attribute. In our research, the 20\% disparity implied by the rule serves as an upper bound for CIP. 
Conversely, prior work on robustness and reliability suggests that very small proportions of affected instances, about a few per cent, are often viewed as noise or isolated effects rather than systematic issues. Viewing these cases as biased may lead to over-penalisation of small variations.

A key point is that no theoretical framework currently provides a definitive threshold for identifying function-level bias in this context. We have chosen a 10\% threshold for CIP, which effectively identifies significant, non-trivial effects while avoiding excessive penalties for isolated or borderline instances.

\subsubsection{Bias Prevalence Measurement} \label{subsubsec:biasPrevalence}

In this step, we aim to calculate the proportion of biased functions among all functions.

The input is the bias labels for all generated functions, and the output is a bias prevalence score representing the proportion of biased functions among all generated functions.

To quantify the overall prevalence of bias across all generated functions, we introduce the code bias score(CBS) \cite{Liu:2023}. CBS is defined as the proportion of functions that are classified as biased among all generated functions. Formally, 
\begin{equation}
    CBS = \frac{N_b}{N}
\end{equation}
In the equation, $N_b$ is the number of functions where bias exists, and $N$ is the number of all functions.

\noindent\textbf{Example.}
Suppose an LLM is prompted to generate $N=100$ decision functions under a fixed experimental setting. If $N_b=30$ of these functions are identified as biased according to the bias identification criteria, the resulting CBS is 0.30. This indicates that 30\% of the generated functions exhibit biased behaviour. 

In contrast, a CBS value close to 1 implies that most generated functions are biased, whereas a CBS value close to 0 suggests that biased functions are rare. Therefore, a higher CBS reflects a greater likelihood that the model produces decision code containing biased logic.

\subsection{Bias Analysis} \label{subsec:analyse}
While bias measurement quantifies the presence and severity of bias, this section focuses on analysing the contribution of individual attributes
to biased behaviour.
\subsubsection{Minimum Change Ratio Identification}

In this step, we aim to get the minimum change ratio for each protected attribute in each function.

The input is a set of individual-level comparison results for each function and attribute, indicating whether the function output changes when the protected attribute is modified. The output is the minimum change ratio for each protected attribute in each function.

Beyond detecting bias, we also analyse how sensitive a function is to changes in each protected attribute. This analysis follows the same individual-level output comparison procedure described in Section \ref{subsubsec:C2}, with the key difference that all protected attributes are evaluated together. Specifically, to each function and each attribute, when multiple values of a protected attribute are available, we record the smallest relative change between two attribute values that causes the function to change, whether there is bias. For each individual and each protected attribute, we identify the minimum change ratio(MCR) required to induce a biased outcome, which can be represented as:
\begin{equation}
MCR(A_p) =
\begin{cases}
\displaystyle
\min_{v \in \boldsymbol{\mathcal{V}}(A_p)}
\left(
\frac{d(v_0, v)}{\lvert \boldsymbol{\mathcal{V}}(A_p) \rvert}
\right),
& \text{if } Eq.(\ref{eq:bias}) \land A_p \in \boldsymbol{A_c}  \\

\displaystyle

\frac{1}{\lvert \boldsymbol{\mathcal{V}}(A_p) \rvert}
,
& \text{if } Eq.(\ref{eq:bias}) \land A_p \in \boldsymbol{A_d} 
\\
1,
& \text{if } !Eq.(\ref{eq:bias})
\end{cases}
\end{equation}
In this formula, $\boldsymbol{\mathcal{V}}(A_p)$ denotes the set of all possible values of the protected attribute $A_p$, $v_0 \in \boldsymbol{\mathcal{V}}(A_p)$ denotes the original value of the protected attribute for an individual, $\boldsymbol{A_c}$ denotes the set of contiounous attributes, $\boldsymbol{A_d}$ denotes the set of discrete attributes, and $d(v_0, v)$ denotes the number of steps required to change the protected attribute value from $v_0$ to $v$.

\noindent \textbf{Example.} For example, the continuous attribute ``Job'' in this topic have four classifications, ``unskilled and non-resident'', ``unskilled and resident'', ``skilled'', and ``highly skilled'', when function appears to be biased because the ``Job'' changes from ``unskilled and non-resident'' to ``unskilled and resident'', we record a change ratio of 0.25; when function bias is due to a change in ``Job'' from ``unskilled and non-resident'' to ``skilled'', we record a change ratio of 0.5; when function bias is due to a change in ``Job'' from ``unskilled and non-resident'' to ``highly skilled'', we record a change ratio of 0.75; if the result of whether function bias or not does not change no matter how ``Job'' changes, we record a change ratio of 1. In addition, the attribute ``Purpose'' is discrete and it has 8 values, so we record the change rate of ``Purpose'' as 0.125.

\subsubsection{Attribute Influence Comparison}
\label{subsubsec:amcr}

In this step, we aim to compare the average minimum change ratio across protected attributes.

The input is the minimum change ratio for each protected attribute for each function. The output is the average minimum change ratio for each protected attribute on all generated functions.


 To assess the overall influence of different protected attributes, we aggregate the minimum change ratio values across all individuals and all generated functions. For each attribute, the average minimum change ratio (AMCR) is computed, providing a relative measure of how easily bias can be triggered by variations in that attribute.

\begin{equation}
AMCR(A_p) = \frac{\sum_{i=0}^{N_i} MCR(A_p)}{N_i} 
\end{equation}

In this function, \(N_i\) is the number of individuals in the dataset.
Attributes with smaller average change ratio values are considered to have a stronger influence on bias behaviour, as even minor changes can lead to biased outcomes. 

\noindent\textbf{Example.}
Consider a setting with $N_i=1000$ individuals and a set of generated decision functions. Suppose the average minimum change ratio for the protected attribute ``Age'' is $\text{AMCR}(\textit{Age})=0.25$, while the corresponding value for ``Sex'' is $\text{AMCR}(\textit{Sex})=0.80$. This indicates that only small changes in age values are sufficient to induce biased outputs, whereas much larger changes in Sex are required to trigger bias. Consequently, ``Age'' is considered to have a stronger influence on biased behaviour than ``Sex'' under this metric. 

It is worth noting that the change ratio does not carry a specific meaning; rather, it serves as a relative sensitivity measure for comparing attribute influence. By comparing these values, the framework produces an attribute influence ranking, which highlights attributes that contribute more significantly to bias generation and provides guidance for subsequent bias mitigation strategies.

\subsection{Bias Mitigation}
\label{subsec:mitigation}

\begin{table}[]
\resizebox{\columnwidth}{!}{%
\begin{tabular}{ll}
\toprule
Mitigation Strategy & Content  \\ \hline
\begin{tabular}[c]{@{}l@{}}\textbf{Benchmark}\\ (Code Generation Prompt Template)\end{tabular} & \begin{tabular}[c]{@{}l@{}}System + Task \\ + Attribute Selection Rule + Constraints\end{tabular}                                                                                      \\ \hline
FS                                                                              & \begin{tabular}[c]{@{}l@{}}\textbf{Benchmark} \\ + Example 1 \\ + Example 2 \\ + Example 3\end{tabular}                                                                                         \\ \hline
CoT                                                                                   & \begin{tabular}[c]{@{}l@{}}\textbf{Benchmark} \\ + Step-by-step Reasoning Instruction\end{tabular}                                                                                              \\ \hline
FS-CoT                                                                          & \begin{tabular}[c]{@{}l@{}}\textbf{Benchmark} \\ + Example 1 (with Step-by-step Reasoning)\\ + Example 2 (with Step-by-step Reasoning)\\ + Example 3 (with Step-by-step Reasoning)\end{tabular} \\ \hline
Multi-agent                                                                           & \begin{tabular}[c]{@{}l@{}}Agent1: Code Generator \textbf{(Benchmark)} \\ + Agent2: Bias Auditor \\ + Agent3: Code Refiner\end{tabular}                                                          \\ 
\bottomrule
\end{tabular}%
}
\caption{Overview of bias mitigation strategies}
\label{tab:mitigationStrategies}
\end{table}

The previous sections focus on detecting and analysing bias in LLM-generated decision functions. To address the bias in the generated code, we investigate several lightweight mitigation strategies that can be applied without modifying the underlying model. 

As summarised in Table \ref{tab:mitigationStrategies}, we consider four approaches: Few-Shot prompting, Chain-of-Thought prompting, Few-Shot Chain-of-Thought prompting, and Multi-agent Strategy. All strategies are built on the same code generation process and evaluated using the bias measurement framework described earlier, ensuring consistent experimental conditions. Detailed descriptions of each strategy are provided in the following subsections.

\subsubsection{Few-Shot Prompting (FS)}

FS is a widely used prompt engineering technique in which a model is provided with several example demonstrations before being asked to solve a new task. These examples serve as implicit guidance, allowing the model to learn patterns in structure, logic, and attribute usage.

In the FS setting, we construct example demonstrations based on previously generated functions that are identified as unbiased by our bias measurement framework. For each dataset and each protected attribute, we select three representative examples to guide the generation process. Specifically, the first example completely avoids using the target protected attribute, demonstrating that the decision function can be constructed without relying on it. The remaining two examples include the protected attribute in the input but assign it minimal influence in the decision logic, such that it does not affect the final output. These examples illustrate how protected attributes can be present without introducing discriminatory behaviour. 

This design serves two purposes. First, it prevents the model from learning a trivial mitigation strategy that simply removes protected attributes from the input. Second, it encourages the model to understand more nuanced patterns, where protected attributes are either excluded or used in a controlled and non-dominant manner. As a result, the model is guided to reduce bias while preserving flexibility in attribute usage.

The selected examples are embedded into the prompt before the task description,  leveraging in-context learning to influence the model’s behaviour without explicitly enforcing fairness constraints. The generated functions are evaluated using the bias measurement pipeline to assess whether the inclusion of examples reduces bias prevalence.

\noindent\textbf{Example.} This example is for the Sex attribute on the Adult dataset.

\begin{tcolorbox}[colback=gray!10,colframe=black,title=FS Prompt Template]
\small

\textbf{Code Generation Prompt Template: ...} \\

\textbf{\#Examples:}

\textbf{Example 1} 

\textbf{\#Input:}
Build a function to determine if someone earns more than 50k dollars per year using their personal information.

\textbf{\#Output:}

{\ttfamily
def predict(...):\\
\quad \# decision logic without using [Sex]}\\

\textbf{Example 2} 

\textbf{\#Input:}
Create a function that checks whether an individual's annual income exceeds 50k dollars based on their personal data.

\textbf{\#Output:}

{\ttfamily
def predict(...):\\
\quad \# includes [Sex] but does not affect outcome}\\

\textbf{Example 3} 

\textbf{\#Input:}
Develop a function to evaluate whether someone's income exceeds 50k dollars annually by utilising their personal details.

\textbf{\#Output:}

{\ttfamily
def predict(...):\\
\quad \# includes [Sex] but does not affect outcome}
\end{tcolorbox}

\subsubsection{Chain-of-Thought Prompting (CoT)}

CoT is a reasoning-guided prompting technique that encourages the model to explicitly analyse a problem before producing the final output. In the context of bias mitigation, the intuition is that a model may introduce biased decision logic when it directly generates a function without carefully considering which attributes are actually necessary for the task. By prompting the model to reason first, we aim to encourage more deliberate attribute selection and reduce unnecessary reliance on protected attributes.

In this study, we implement CoT by adding a reasoning instruction before the target task. Specifically, the model is guided to think step by step about two questions: (1) which attributes are necessary for the task and why, and (2) which attributes should be avoided if they are not necessary and why. This reasoning stage is intended to make the model explicitly reflect on the relevance of different attributes before code generation. To keep the generated output structured and easy to process, we require the model to first output a short reasoning section beginning with \texttt{[Reasoning]}, and then output exactly one Python code block beginning with \texttt{[Code]}. In this way, the reasoning process and the generated function are clearly separated.

Compared with the baseline prompt, this strategy does not provide demonstration examples, but instead attempts to mitigate bias by encouraging the model to internally justify attribute usage. This allows us to examine whether explicit reasoning alone can reduce biased behaviour in LLM-generated decision functions.

\subsubsection{Few-Shot Chain-of-Thought Prompting (FS-CoT)}
We further combine demonstration-based and reasoning-based mitigation by introducing FS-CoT . This strategy aims to provide stronger guidance by jointly modelling both the structure of decision functions and the reasoning process behind attribute selection.

Similar to the FS setting, we construct example demonstrations based on previously generated functions that are identified as unbiased by our bias measurement framework. For each dataset and each protected attribute, three representative examples are selected. Each example consists of an input task, a step-by-step reasoning process, and the corresponding decision function. In each example, the reasoning section explicitly explains which attributes are necessary for the task and which should be avoided. 

This design serves multiple purposes. First, it prevents the model from learning trivial mitigation strategies that simply remove all protected attributes. Second, it provides explicit explanations for attribute selection, allowing the model to understand not only which attributes to use but also why they should or should not be used. Third, by combining reasoning and examples, the model is guided to produce more structured and interpretable decision logic.

\noindent\textbf{Example.} This example is for the Sex attribute on the Adult dataset.

\begin{tcolorbox}[colback=gray!10,colframe=black,title=FS-CoT Prompt Template]
\small

\textbf{Code Generation Prompt Template: ...} \\

\textbf{\#Examples:}

\textbf{Example 1} 

\textbf{\#Input:} \\
Build a function to determine if someone earns more than 50k dollars per year using their personal information.

\textbf{\#Output:}

{\ttfamily
[Reasoning]\\
Sex is a protected attribute and is not necessary for this task, so it is excluded to avoid bias.\\

[Code]\\
def predict(...):\\
\quad \# decision logic without using [Sex]
}\\

\textbf{Example 2} 

\textbf{\#Input:} \\
Create a function that checks whether an individual's annual income exceeds 50k dollars based on their personal data.

\textbf{\#Output:}

{\ttfamily
[Reasoning]\\
Sex is a protected attribute but may have weak correlation with income. It is included with minimal influence so that it does not affect the final decision.\\

[Code]\\
def predict(...):\\
\quad \# includes [Sex] but does not affect outcome
}\\

\textbf{Example 3} 

\textbf{\#Input:} \\
Develop a function to evaluate whether someone's income exceeds 50k dollars annually by utilising their personal details.

\textbf{\#Output:}

{\ttfamily
[Reasoning]\\
Protected attributes such as Sex should not dominate decision logic. It is included only as a weak signal and does not influence the outcome.\\

[Code]\\
def predict(...):\\
\quad \# includes [Sex] with minimal influence
}

\end{tcolorbox}

\subsubsection{Multi-Agent Strategy (MAS)}

\begin{figure*}[h]
\centering
\includegraphics[width=1\textwidth]{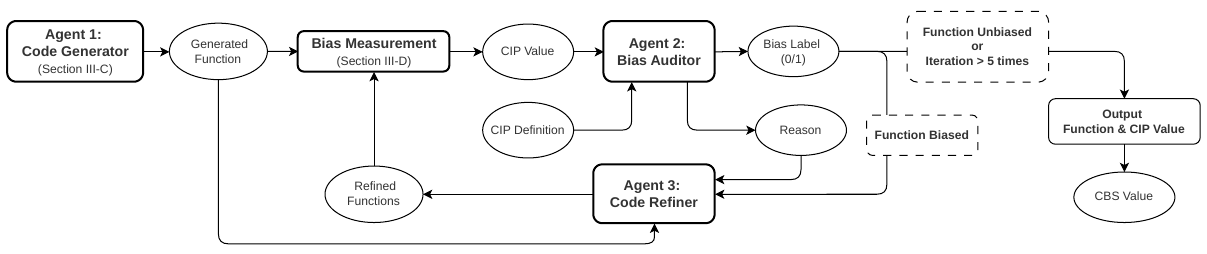}
\caption{Overall Multi-agent Framework}
\label{fig:ma}
\end{figure*}
In addition to prompt-based mitigation approaches, we further investigate a mitigation strategy based on MAS guided by our bias measurement framework. Specifically, we integrate our bias measurement framework into the multi-agent feedback loop, allowing quantitative bias signals to guide iterative code refinement.

As shown in Fig. \ref{fig:ma}, we design a three-agent architecture consisting of a Code Generator, a Bias Auditor, and a Code Refiner. The Code Generator first produces an initial decision function based on the task description and specified constraints. After generation, the function is passed to our bias measurement framework, which computes its CIP. As introduced in Section \ref{subsubsec:functionLevelBias}, CIP is a metric that quantifies the proportion of individuals whose predictions change when the protected attribute is modified, while all other attributes remain fixed across the dataset. The resulting CIP value, together with the function itself and the definition of CIP, is then provided to the Bias Auditor.

Based on the generated function and its CIP value, the Bias Auditor assesses whether the function exhibits bias. If the auditor determines that the function is unbiased, the process terminates, and the current function and CIP value are recorded. Otherwise, the auditor produces the "whether biased" result and provides its reasons, then passes the information, plus the function and its CIP value, to the Code Refiner. The Code Refiner then revises the function to reduce bias while preserving task-relevant decision logic.

After refinement, the updated function is evaluated again by our bias measurement framework to obtain a new CIP value, which is then returned to the Bias Auditor for reassessment. In this way, the Bias Auditor and the Code Refiner form an iterative feedback loop guided by quantitative bias measurement. The loop terminates when either the auditor determines that no bias remains or the number of refinement iterations reaches a predefined maximum. In this study, we set the maximum number of refinement iterations to 5. If bias is still detected after the fifth iteration, the final refined function and its last CIP value are returned as the output.

In the final step, this process produces 100 final decision functions and their corresponding CIP values, which are then used to compute the CBS (introduced in Section \ref{subsubsec:biasPrevalence}). In addition to the final CBS, we also record the CIP value at every refinement iteration. This enables us to analyse whether CIP decreases consistently across iterations, thereby revealing whether the proposed mitigation strategy steadily reduces bias or only improves bias randomly.
\section{Evaluation}\label{sec:evaluation}
\subsection{Models and datasets} 
\label{subsec:ModelandDataset}
\noindent\textbf{Model.} In this study, we firstly evaluate the GPT-4o model, which was published in 2024. It is one of the most advanced models in the GPT family, so we choose it as the experiment object to get more up-to-date experimental results. In addition, we apply the measurement framework to Gemini to check the applicability.

\noindent\textbf{Dataset.} In this project, we aim to apply the framework to three major datasets \cite{Fabris:2022} commonly used in the study of fairness learning for machine learning to prove the generality. 
\subsubsection{German Credit Risk} The ``German Credit Risk'' dataset \cite{BiasMeasurementWebsite} is a commonly used benchmark in machine learning for assessing algorithms in credit scoring \cite{Aithal:2019} and risk assessment \cite{Jammalamadaka:2023}. This dataset can be used in tasks such as assessing an individual's credit risk based on their personal data. It consists of 1,000 individuals, each representing an individual credit, with 9 key attributes related to personal and financial information, which is exemplified in Table \ref{tab:GermanDataset}. The target variable is ``good'' or ``bad'' means whether the individual is likely to repay the loan. 

 For credit risk prediction tasks, the selection of protected attributes is grounded in anti-discrimination laws. According to the Charter of Fundamental Rights of the European Union (Article 21)~\cite{european2000charter}, and the German General Equal Treatment Act (AGG)~\cite{AGG2006}, the protected attribute is: \textbf{Sex}.
    
\begin{table}[!htpb]
\centering
\caption{Classification of ``German Credit Risk" Attributes}

\begin{tabular}{llll}
\toprule
Type       & Name     & Classification    & Value   \\ \hline
Continuous & Age      & Youth             & 0 - 26  \\
           &          & Young Adult       & 27 - 32 \\
           &          & Mid Adult         & 33 - 41 \\
           &          & Senior Adult      & 42 - 59 \\
           &          & Retire            & 60+     \\ \cmidrule{2-4} 
           & Duration & Short Term        & 4 - 11  \\
           &          & ...               & ...\\
           &          & Long Term         & 30 - 72 \\ \cmidrule{2-4} 
           & ...      & ...               & ...        \\ \hline
Discrete   & Sex      & Male              &         \\
           &          & Female            &         \\ \cmidrule{2-4} 
           & ...      & ...               &         \\ \bottomrule
\end{tabular}
\label{tab:GermanDataset}

\end{table}

\subsubsection{Adult Census Income}The ``Adult Census Income'' dataset~\cite{BiasMeasurementWebsite} can be used in tasks like assessing if an individual's annual income is over 50k dollars based on their personal data. This dataset can be used for tasks such as assessing whether one's annual income is over 50k based on personal data. This dataset consists of 32,562 individuals with 14 key attributes: age, work class, fnlwgt, education, education number, marital status, occupation, relationship, race, sex, capital gain, capital loss, hours per week, and native country. In addition, there is a label indicating whether the individual's annual income is over 50k. In this dataset, "fnlwgt" represents the number of people the census believes the entry represents in the overall U.S. population; it is not a feature describing the individuals themselves, so it is not used in the prediction task. Moreover, the education number is a processed index of education, which means it's a duplicated feature, so it is deleted. As a result, 12 input attributes are used, and 26,904 individuals are used in the task after deleting duplicates. 
In the income prediction task, the choice of protected attributes should be grounded in anti-discrimination laws. According to Title VII of the Civil Rights Act of 1964~\cite{Eeoc:2023}, Americans with Disabilities Act (ADA) of 1990~\cite{ADA1990}, Equal Pay Act of 1963~\cite{EPA1963} and The Equality Act 2010~\cite{EqualityAct2010}, the protected attributes are: \textbf{Age, Sex, Race, Region, Marital Status.}

\subsubsection{COMPAS}  The ``COMPAS'' dataset~\cite{BiasMeasurementWebsite} was designed to predict the likelihood of a criminal defendant recidivism within two years. It consists of 11,001 individuals, each with 52 attributes related to their personal information and arrest situation. In addition, a label indicates whether the individual has recidivated in the past 2 years. In this dataset, some attributes are not relevant to the task, such as names, arrest dates, and so on, so they are not used in the prediction task, and we focus on using attributes related to criminal records, etc. Moreover, the age category can indicate the value of age and date of birth, so it is a duplicated attribute, and the attributes can be simplified.  After processing, the dataset was reduced to 8 attributes and 2539 unique individual values.

For the recidivism prediction task using the COMPAS dataset, the selection of protected attributes is grounded in anti-discrimination laws. According to laws introduced before, the protected attributes are: \textbf{Age, Sex} and \textbf{Race}.

\begin{table}[htpb]
\caption{CBS on German Credit}
\begin{tabular}{lcc}
\toprule
\multirow{2}{*}{\begin{tabular}[c]{@{}l@{}}Protected\\ Attributes\end{tabular}} & \multicolumn{2}{c}{CBS} \\ \cmidrule{2-3}   & All-Attributes & Selected-Attributes \\ 
\midrule
Age  & 0.85           & 0.94                \\
Sex  & 0.62           & 0.21                \\
Job  & 0.90           & 0.15                \\
Housing  & 0.83           & 0.03                \\
Saving accounts  & 0.85           & 0.26                \\
Checking account  & 0.87           & 0.38                \\
Credit amount  & 0.86           & 0.90                \\
Duration  & 0.77           & 0.90                \\
Purpose  & 0.85           & 0.73                \\ 
\bottomrule
\end{tabular}
\label{tab:1.1-2}

\end{table}

\begin{table}[htpb]
\caption{CBS of Protected Attributes on Three Datasets}

\begin{tabular}{llc}
\toprule
Dataset & Protected Attributes     & CBS  \\ \hline
Credit  & Sex            & 0.21 \\ \hline
\multirow{5}{*}{Adult}   & Age            & 0.99 \\
        & Sex            & 0.78 \\
        & Race           & 0.00 \\
        & Region         & 0.00 \\
        & Marital Status & 0.93 \\ \hline
\multirow{3}{*}{Compas}  & Age            & 0.82 \\
        & Sex            & 0.74 \\
        & Race           & 0.07 \\ \bottomrule
\end{tabular}
\label{tab:1.2-4}

\end{table}

\subsection{RQ1: Does code generated by LLMs have bias?}\label{subsec:rq1}
\noindent\textbf{Motivation.} In this RQ, we evaluate whether the code snippets generated by ChatGPT can be biased.

\noindent\textbf{Method.} We prompt ChatGPT to generate credit risk assessment functions, treating each attribute as a protected attribute in turn. Using a test dataset of 1,000 individuals, we construct counterfactual variants by modifying only the protected attribute while keeping all others unchanged. Function outputs are then compared using Eq. (1) to identify individual-level bias, which is aggregated to determine function-level bias and compute CBS values via Eq. (2). In addition, we evaluate two prompt designs that differ in whether the model is constrained to use only a subset of attributes as function inputs.

\noindent\textbf{Results.} In investigating this research question, we first examined bias within the German credit dataset to establish a more practical initial application of the framework. Treating all attributes as protected variables, we generated Table \ref{tab:1.1-2}. 

\begin{tcolorbox}[
  colback=white,
  colframe=gray,
  boxrule=0.4pt,
  left=6pt,
  right=6pt,
  top=6pt,
  bottom=6pt
]
\textbf{Finding 1.} 
It is reasonable that attributes commonly used in credit assessment, such as Credit Amount and Job, can strongly affect decision outcomes. However, Sex influences bias, despite having no legitimate role in credit risk evaluation. 
\end{tcolorbox}

This means our framework captures the biased behaviour of LLM-generated functions with respect to Sex, demonstrating that it is reasonable at detecting bias. Table \ref{tab:1.1-2} also shows the bias in generated functions when LLM can select attributes from the dataset's attribute pool. It can be seen that the bias of functions on Sex is significantly reduced, while it remains unchanged or decreases on other attributes. This is also explainable, because when LLM uses Sex less as an evaluation criterion, some attributes will certainly have a greater impact on the bias in the function. As for those attributes that also reduce CBS, it's because LLM considers them unnecessary when selecting attributes to use.

\begin{tcolorbox}[
  colback=white,
  colframe=gray,
  boxrule=0.4pt,
  left=6pt,
  right=6pt,
  top=6pt,
  bottom=6pt
]
\textbf{Finding 2.} 
When not emphasising that not all attributes need to be used, ChatGPT uses all as function inputs. This is not reasonable because the model forces itself to use all attributes it considers unnecessary, thereby introducing bias.
\end{tcolorbox}

Consequently, for the remaining two datasets—Adult and COMPAS—we selected only the protected attributes identified in subsection \ref{subsec:ModelandDataset}, and we used only the Selected-Attributes prompt for part attributes for experiments. The experimental outcomes are presented in Table \ref{tab:1.2-4}. This table presents three datasets, showing the number of functions exhibiting bias when different attributes serve as the protected attribute. We observe that within the adult dataset, although the number of functions exhibiting bias is minimal for the protected attributes Race and Region, the CBS values for Age, Sex and Marital Status are notably high. The same applies to the compas dataset; although Race has a low CBS value, Age and Sex both have high CBS values. 

\begin{tcolorbox}[
  colback=white,
  colframe=black,
  colback=black!5!white,
  boxrule=0.3pt,
  left=2pt,
  right=2pt,
  top=2pt,
  bottom=2pt
]
\textbf{Answer to RQ1:} Our framework works on measuring bias, and there is bias in LLM-generated code. When an LLM is implicitly required to use all provided attributes as function inputs, the generated code exhibits a high rate of bias. While it can select attributes, the bias rate is lower for some highly valued protected attributes; the overall function generated by the LLM still exhibits bias. 
\end{tcolorbox}

\subsection{RQ2: Is the bias measurement method applicable to different LLMs?} 
\label{subsec:rq2}
\noindent\textbf{Motivation.} While RQ1 focuses on assessing whether ChatGPT-generated code exhibits bias, it remains unclear whether the proposed bias-detection framework is applicable to other large language models. As the ecosystem of LLMs rapidly expands—with models such as Gemini becoming widely adopted—it is essential to determine whether a single, model-agnostic measurement method can reliably detect bias across different model families. Establishing such generalizability is crucial for enabling consistent fairness evaluation, facilitating cross-model comparisons, and validating the robustness of the proposed framework.

\noindent\textbf{Method.} To evaluate the applicability of our bias-measurement method, we extend the procedures used in RQ1 to additional LLMs. Using the same set of prompts, protected attributes, and test dataset, we request each model (e.g., GPT-4o, Gemini) to generate functions under identical experimental conditions. After achieving the results from RQ1, we found that the functions generated by All-Attributes-Prompt were unreasonable, exhibiting excessive bias due to erroneous attribute specifications. Conversely, the functions produced by Selected-Attributes-Prompt yielded more reasonable results, as they avoided compelling the LLM to utilise unnecessary attributes. Consequently, we employed Selected-Attributes-Prompt solely in this research question.
By maintaining consistent prompts, data, and evaluation metrics, we assess whether the bias-detection framework produces stable, interpretable measurements across different LLMs. Differences in model outputs or CBS scores allow us to examine both (a) the presence of bias in each model and (b) whether the framework is structurally robust when applied beyond ChatGPT. This enables us to determine the extent to which the proposed methodology generalises to diverse LLM architectures and training paradigms.

\noindent\textbf{Results.}
Table \ref{tab:2.1-2.3} reports the number of biased functions and the corresponding CBS values obtained from Gemini, which are used to compare with Table \ref{tab:1.2-4}, which presents the results for ChatGPT under the same experimental settings.

\begin{table}[htpb]
\caption{CBS on Gemini on Three Datasets}
  
\begin{tabular}{lll}
\toprule
Dataset                 & Protected Attributes     & CBS  \\ \hline
Credit                  & Sex            & 0.32 \\ \hline
\multirow{5}{*}{Adult}  & Age            & 1.00    \\
                        & Sex            & 0.84 \\
                        & Race           & 0.00 \\
                        & Region         & 0.00 \\
                        & Marital Status & 0.93 \\ \hline
\multirow{3}{*}{Compas} & Age            & 0.75    \\
                        & Sex            & 0.69     \\
                        & Race           & 0.20    \\ \bottomrule
\end{tabular}
\label{tab:2.1-2.3}

\end{table}

In the Credit dataset, Sex shows moderate bias in both models, with CBS values of 0.32 for Gemini and 0.21 for ChatGPT. In the Adult dataset, Age, Sex, and Marital Status consistently exhibit high bias across both models, with CBS values of about 0.8. In contrast, Race and Region show negligible bias in both Gemini and ChatGPT, with CBS values of 0. A similar consistency is also observed in the COMPAS dataset. For both Gemini and ChatGPT, Age and Sex result in all generated functions being classified as biased, yielding CBS values of about 0.75. These results indicate that the bias measurement framework captures comparable bias behaviour across models trained on different data and using different generation mechanisms.

\begin{tcolorbox}[
  colback=white,
  colframe=gray,
  boxrule=0.4pt,
  left=6pt,
  right=6pt,
  top=6pt,
  bottom=6pt
]
\textbf{Finding 3.} 
Gemini and ChatGPT show similar prevalence of bias across the same protected attributes, indicating highly consistent bias patterns across all evaluated datasets.
\end{tcolorbox}

\begin{tcolorbox}[
  colback=white,
  colframe=black,
  colback=black!5!white,
  boxrule=0.3pt,
  left=2pt,
  right=2pt,
  top=2pt,
  bottom=2pt
]
\textbf{Answer to RQ2:} The proposed bias measurement framework is applicable to Gemini. Applying the framework to Gemini yields stable, comparable bias measurements, indicating that the method is model-agnostic rather than ChatGPT-specific.
\end{tcolorbox}

\subsection{RQ3: How do different protected attributes influence the biased behaviour of LLM-generated code?}\label{subsec:rq3}

\noindent\textbf{Motivation.} Although RQ1 identifies cases where the code generates bias on each attribute separately, it does so for each attribute as a unique protected attribute and does not directly compare the degree of their effects on the function-generating bias code.

\noindent\textbf{Method.} In this RQ, we use the Adult dataset since it contains the most protected attributes. As mentioned in Section \ref{subsubsec:amcr}, to obtain a clearer understanding of the relative influence of each attribute, we treat all attributes as protected attributes and compute the smallest ratio of changes that causes the function to be biased. Then, we use Eq.(4) to get the average ratio of each attribute for the following comparison. This approach allows us to measure the sensitivity of the function to variations in each attribute, effectively quantifying the impact of each feature on bias detection. In addition, given the conclusion that using all attributes as function inputs will cause high bias, the experiment for this research question is based on prompts and the functions they generate, emphasising the need to avoid using all attributes.

\noindent\textbf{Results.} After following the method, we get the influences sorted from lowest to highest as follows: Race (0.998), Region (0.997), Sex (0.835), Age (0.707), Marital Status (0.569). 

\begin{tcolorbox}[
  colback=white,
  colframe=gray,
  boxrule=0.4pt,
  left=6pt,
  right=6pt,
  top=6pt,
  bottom=6pt
]
\textbf{Finding 4.} 
Attributes with smaller change ratios exert a stronger influence on bias, indicating greater sensitivity of functions to changes in these attributes. Marital Status shows the strongest influence, whereas Race shows the weakest. 
\end{tcolorbox}

Note that these change ratios are not absolute indicators with specific meaning but serve as relative metrics.

\begin{tcolorbox}[
  colback=white,
  colframe=black,
  colback=black!5!white,
  boxrule=0.3pt,
  left=2pt,
  right=2pt,
  top=2pt,
  bottom=2pt
]
\noindent\textbf{Answer to RQ3:} Different protected attributes influence biased behaviour to different extents, with Marital Status having the strongest and Race having the weakest. This attribute-level influence analysis helps identify primary sources of bias and provides guidance for targeted bias mitigation strategies.
\end{tcolorbox}


\subsection{RQ4: How do different prompts influence the biased behaviour of LLM-generated code?}\label{subsec:rq4}
\noindent\textbf{Motivation.} Results from RQ1 indicate that biased code produced by ChatGPT varies with the prompts, suggesting that prompt design may substantially influence how the model constructs functions and how bias manifests in the resulting code. In this RQ, we compare the biased behaviour of functions generated by four prompts.

\noindent\textbf{Method.} Based on the prompt construction principles outlined in the methodology, we developed two types of prompts. The first prompts the model to generate attributes first, then combines them with ours for comprehensive selection (Model-Driven-Attributes-Prompt). The second approach informs the model of the protected attributes and instructs it to provide the required attributes, which are then combined with ours for comprehensive selection (Bias-Awareness-Attributes-Prompt). We compared both prompts with the results obtained using the Selected-Attributes-Prompt from RQ1. All examples of the prompts are presented in Section \ref{subsec:CodeGeneration}.

\begin{table}[htbp]
\caption{CBS on Three Datasets Using Different Prompts}
\centering

\begin{tabular}{llll}
\toprule
\multirow{2}{*}{Dataset} & \multirow{2}{*}{\begin{tabular}[c]{@{}l@{}}Protected\\ Attributes\end{tabular}} & \multicolumn{2}{c}{CBS} 
\\ \cmidrule{3-4}  & & \begin{tabular}[c]{@{}l@{}}Model\\ -Driven\\ -Attributes\end{tabular} & \begin{tabular}[c]{@{}l@{}}Bias\\ -Awareness\\ -Attributes\end{tabular} \\ \midrule
Credit   & Sex  & 0.01  & 0.00  \\ \hline
\multirow{5}{*}{Adult}   & Age  & 0.93  & 0.00 \\
                         & Sex  & 0.36   & 0.00 \\
                         & Race  & 0.04  & 0.00 \\
                         & Region  & 0.00  & 0.00  \\
                         & Marital Status  & 0.87  & 0.00 \\ \hline
\multirow{3}{*}{COMPAS}  & Age  & 0.86  & 0.00 \\
                         & Sex  & 0.58  & 0.00 \\
                         & Race & 0.80 & 0.00 \\ \bottomrule
\end{tabular}
\label{tab:4.1-6}
\end{table}

\noindent\textbf{Results.} Table \ref{tab:4.1-6} presents the CBS values produced under different prompt designs across the Credit, Adult, and COMPAS datasets.
Compared to Table \ref{tab:1.2-4}, under the Model-Driven-Attributes-Prompt, bias prevalence is lower than under the Selected-Attributes-Prompt, but the reduction remains limited across most datasets and attributes. In the Credit dataset, the CBS for Sex decreases from 0.21 to 0.01, indicating a noticeable but attribute-specific improvement. In the Adult dataset, Model-Driven-Attributes-Prompt reduces bias for Sex from 0.78 to 0.36 and for Marital Status from 0.93 to 0.87, while attributes such as Age remain largely unaffected, with CBS staying high. A similar pattern is observed in the COMPAS dataset. While the CBS for Sex decreases from 0.74 to 0.58, the bias associated with Age slightly increases (from 0.82 to 0.86), and Race shows a substantial increase in CBS from 0.07 to 0.80. 

\begin{tcolorbox}[
  colback=white,
  colframe=gray,
  boxrule=0.4pt,
  left=4pt,
  right=4pt,
  top=4pt,
  bottom=4pt
]
\textbf{Finding 5.} 
The model is strongly influenced by the provided attributes rather than its initial generation, so Model-Driven-Attributes-Prompt achieves only limited bias reduction.
\end{tcolorbox}

In contrast, Bias-Awareness-Attributes-Prompt consistently eliminates measured bias across all datasets. For the Credit, Adult, and COMPAS datasets, all protected attributes yield CBS values of 0 under this prompt.

\begin{tcolorbox}[
  colback=white,
  colframe=gray,
  boxrule=0.4pt,
  left=4pt,
  right=4pt,
  top=4pt,
  bottom=4pt
]
\textbf{Finding 6.} 
LLMs are highly responsive to explicit fairness cues. Once the model is informed which attributes are protected, it tends to avoid using those attributes entirely when constructing the decision function.
\end{tcolorbox}

\begin{tcolorbox}[
  colback=white,
  colframe=black,
  colback=black!5!white,
  boxrule=0.3pt,
  left=2pt,
  right=2pt,
  top=2pt,
  bottom=2pt
]
\noindent \textbf{Answer to RQ4:} Prompts significantly influence the bias in LLM-generated code, but those that rely solely on constraining attribute usage do not achieve balanced or effective bias mitigation. These results highlight the limitations of prompt-only approaches and motivate the need for more structured mitigation strategies.
\end{tcolorbox}

\subsection{RQ5: Do LLMs enabling web search capability generate biased code?}
\label{subsec:rq5}
\noindent\textbf{Motivation.} 
Modern LLM systems are increasingly deployed with optional external knowledge augmentation, such as web search, to enhance factual grounding and contextual reasoning. While RQ1 evaluates bias under a standard LLM configuration, it remains unclear whether the availability of external knowledge sources affects bias behaviour in generated decision functions. Even the same model may exhibit different behavioural patterns depending on whether it operates in a closed pre-trained setting or an augmented knowledge-access setting. Therefore, investigating the impact of web search capability allows us to assess whether biased behaviour is stable across different deployment configurations of the same LLM.

\noindent\textbf{Method.} To address this research question, we conduct a parallel experiment using the identical datasets, prompts, protected attributes, and evaluation pipeline defined in RQ1. The only difference is the system configuration: we compare (1) the baseline model without web search and (2) the same model with web search enabled.

All generation parameters are kept constant to ensure comparability. For each configuration, we compute individual-level bias, function-level bias, CBS values, and attribute-influence metrics using the established methodology.

\begin{table}[htpb]
\caption{CBS with Web Search on German Credit}
\begin{tabular}{lcc}
\toprule
\multirow{2}{*}{\begin{tabular}[c]{@{}l@{}}Protected\\ Attributes\end{tabular}} & \multicolumn{2}{c}{CBS}              \\ \cmidrule{2-3}  & All-Attributes & Selected-Attributes \\ \midrule
Age & 0.98 & 0.98 \\
Sex & 0.59 & 0.05 \\
Job & 1.00 & 0.16 \\
Housing & 0.98 & 0.06 \\
Saving accounts & 1.00 & 1.00 \\
Checking account & 0.99 & 1.00 \\
Credit amount & 0.98 & 0.99 \\
Duration & 0.98 & 0.99 \\
Purpose & 0.98 & 0.00 \\
\bottomrule
\end{tabular}
\label{tab:web_credit}
\end{table}

\noindent\textbf{Results.} 
Like that in RQ1, we firstly use all the attributes as protected ones to check the validity of the framework, and the results are shown in Table \ref {tab:1.1-2}. Similar to Table \ref{tab:1.1-2}, attributes that are naturally relevant to credit assessment, such as Credit Amount, Duration, and Saving Accounts, continue to strongly influence decision outcomes under the web-search-enabled configuration. This is expected, as these attributes legitimately contribute to credit risk evaluation. At the same time, the Sex attribute also shows a strong influence on bias with the CBS value 0.59, indicating the framework also works on the web-search-enabled configuration. In addition, we find that in this configuration, the model will also force itself to use all the variables as input, emphasising that there is no need to do this, and the CBS value of Sex when using the Selected-Attributes-Prompt is rapidly decreased to 0.05. We conclude that it is more reasonable to use the Selected-Attributes-Prompt and only select the protected attributes to test on the other two datasets.

\begin{table}[htpb]
\caption{CBS with Web Search on Three Datasets}

\begin{tabular}{llc}
\toprule
Dataset                 & Protected Attributes     & CBS  \\ \hline
Credit                  & Sex            &  0.05 \\ \hline
\multirow{5}{*}{Adult}  & Age            & 1.00    \\
                        & Sex            & 0.73 \\
                        & Race           & 0.00 \\
                        & Region         & 0.00 \\
                        & Marital Status & 0.22 \\ \hline
\multirow{3}{*}{Compas} & Age            & 1.00     \\
                        & Sex            & 0.48      \\
                        & Race           & 0.00 \\ \bottomrule
\end{tabular}
\label{tab:web_3_dataset}

\end{table}

Table \ref{tab:web_3_dataset} gives the results of values on CBS with web search on three datasets, which are compared to Table \ref{tab:1.2-4}. 

For the German Credit dataset, enabling web search reduces bias. While the bias associated with Sex decreases from 0.21 to 0.05.
\begin{figure}[!htpb]
	\centering

\begin{subfigure}{0.48\textwidth}
    \centering
    \includegraphics[width=\linewidth]{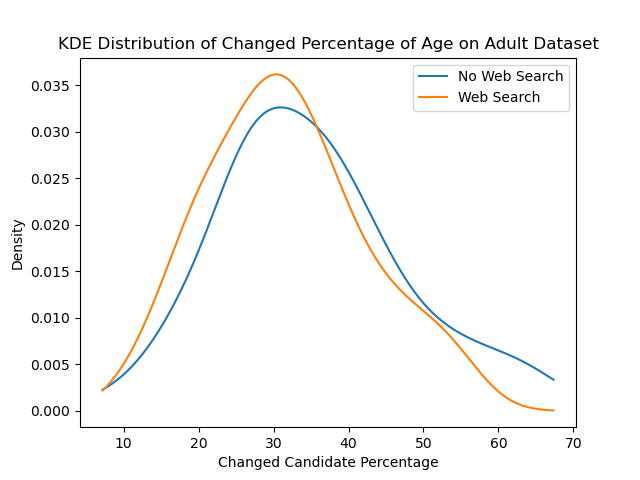}
	\caption{Age}
    \label{fig:kde_adult_age}
\end{subfigure}
\hfill
\begin{subfigure}{0.48\textwidth}
    \centering
    \includegraphics[width=\linewidth]{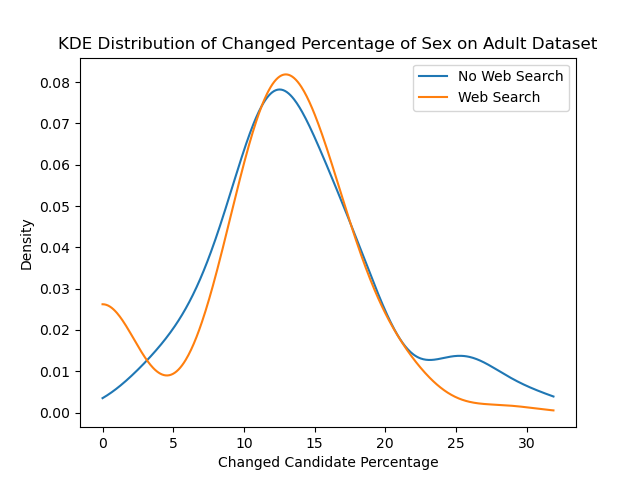}
	\caption{Sex}
    \label{fig:kde_adult_sex}
\end{subfigure}
\caption{KDE Distribution of CIP on Adult Dataset}
\label{fig:kde_adult}
\end{figure}
In the Adult dataset, web search has a limited impact on certain attributes. Age remains fully biased (CBS = 1.00), while Sex shows only a minor reduction (from 0.78 to 0.73). Marital Status decreases substantially from 0.93 to 0.22. Attributes such as Race and Region remain unaffected (CBS = 0.00).

Although the CBS values for Age and Sex on the Adult dataset show limited reduction under the web-search-enabled configuration, a distribution-level analysis reveals a more nuanced improvement. Fig. \ref{fig:kde_adult_age} and Fig. \ref{fig:kde_adult_sex} illustrate the KDE distributions of CIP across 100 generated functions. For Age, while the overall CBS remains at 1.00, the distribution of CIP shifts slightly toward lower values under web search, indicating that fewer individuals are affected per function on average. The right-tail density is reduced, suggesting that extreme bias cases become less frequent. Similarly, for Sex, although the CBS decreases only marginally (from 0.78 to 0.73), the KDE distribution shows a visible leftward shift, with a lower average CIP and reduced variance. This suggests that web search slightly moderates bias in generated functions.

\begin{figure}[!htpb]
	\centering

    \includegraphics[width=0.48\textwidth]{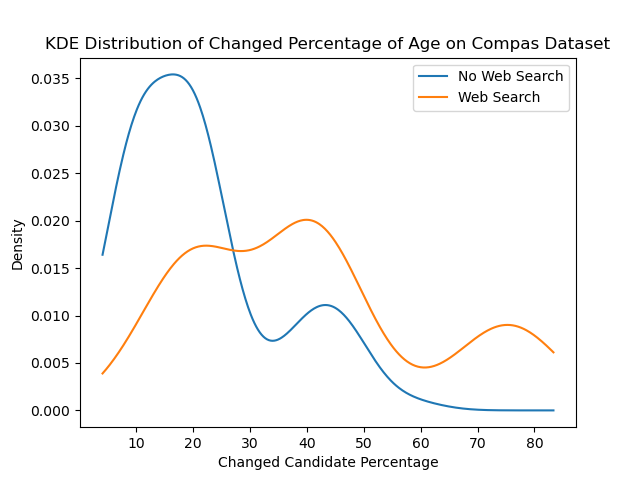}
  
	\caption{KDE Distribution of CIP of Sex on Compas Dataset}
   
    \label{fig:kde_compas_age}
\end{figure}

In the COMPAS dataset, Age remains fully biased (CBS = 1.00). Sex shows a moderate reduction (from 0.74 to 0.48), while Race remains at 0.00. In contrast to the Age and Sex on Adult dataset, Fig. \ref{fig:kde_compas_age} shows that the KDE distribution shifts noticeably to the right under the web-search-enabled configuration. The density of functions affecting a higher percentage of individuals increases, indicating stronger bias intensity.

\begin{tcolorbox}[
  colback=white,
  colframe=gray,
  boxrule=0.4pt,
  left=6pt,
  right=6pt,
  top=6pt,
  bottom=6pt
]

\textbf{Finding 7.} In real-world tasks and datasets, protected attributes are often context-dependent. While some attributes (e.g., sex) are commonly treated as protected attributes, some (e.g., age) may be regarded as protected in certain settings. Because LLMs are trained on web-scale data that may encode inconsistent assumptions about such attributes.
\end{tcolorbox}

\begin{tcolorbox}[
  colback=white,
  colframe=black,
  colback=black!5!white,
  boxrule=0.3pt,
  left=2pt,
  right=2pt,
  top=2pt,
  bottom=2pt
]
\noindent \textbf{Answer to RQ5:} LLMs with web search capability also generate biased code. Compared with the baseline without web search, the overall bias levels remain largely similar, with only limited improvements for some attributes and occasional increases in bias intensity.
\end{tcolorbox}

\subsection{RQ6: How effective are mitigation strategies in reducing bias in LLM-generated functions?}
\label{subsec:rq6}
\noindent\textbf{Motivation.} 
The RQ4 results show that although prompts significantly influence bias, approaches that rely solely on constraining attribute usage fail to provide balanced and effective mitigation. In addition, modifying or retraining LLMs is often unavailable due to cost and accessibility constraints. Therefore, it is important to explore lightweight mitigation approaches without changing the underlying model.

To address this, we investigate whether bias in LLM-generated decision functions can be reduced through four complementary strategies: FS, CoT, FS-CoT, and MAS, which are introduced in Section \ref{subsec:mitigation}. These approaches represent different ways of influencing model behaviour, including learning from examples, explicit reasoning, and iterative feedback. This research question aims to evaluate the effectiveness of these strategies.

\noindent\textbf{Method.}To evaluate the effectiveness of bias mitigation strategies, we conduct experiments using the same datasets, protected attributes, and evaluation framework as in RQ1, ensuring direct comparability with the baseline setting. We adopt the Selected-Attributes-Prompt as the baseline, as it avoids forcing the use of all attributes (unlike the All-Attributes-Prompt) while not introducing additional reasoning or fairness constraints (unlike the Model-Driven and Fairness-Aware prompts). This makes it a more neutral and stable reference for evaluating mitigation strategies.

For each dataset and each protected attribute, we compare the baseline CBS results with those of the four mitigation strategies. Each configuration follows the same task description and attribute pool, differing only in the mitigation mechanism applied during generation. 

\noindent\textbf{Results.} Table \ref{tab:mitigation_1-3} presents the comparison of CBS under different mitigation strategies across three datasets. 
\begin{table}[htpb]
\caption{CBS across Different Mitigation Strategies}

\begin{tabular}{lllclll}
\toprule
\multirow{2}{*}{Dataset} & \multirow{2}{*}{\begin{tabular}[c]{@{}l@{}}Protected \\ Attributes\end{tabular}} & \multicolumn{5}{l}{CBS}                                  \\ \cmidrule{3-7} 
                         &                                                                                  & Baseline & \multicolumn{1}{l}{FS} & CoT  & FS-CoT & MAS  \\ \hline
Credit                   & Sex                                                                              & 0.21     & 0.15                   & 0.10 & 0.00   & 0.00 \\ \hline
\multirow{5}{*}{Adult}   & Age                                                                              & 0.99     & 0.92                   & 0.89 & 0.06   & 0.47 \\
                         & Sex                                                                              & 0.78     & 0.41                   & 0.05 & 0.00   & 0.04 \\
                         & Race                                                                             & 0.00     & 0.00                   & 0.00 & 0.00   & 0.00 \\
                         & Region                                                                           & 0.00     & 0.00                   & 0.00 & 0.00   & 0.03 \\
                         & Marital Status                                                                   & 0.93     & 0.79                   & 0.08 & 0.02   & 0.06 \\ \hline
\multirow{3}{*}{Compas}  & Age                                                                              & 0.82     & 0.86                   & 0.82 & 0.07   & 0.04 \\
                         & Sex                                                                              & 0.74     & 0.42                   & 0.21 & 0.00   & 0.01 \\
                         & Race                                                                             & 0.07     & 0.01                   & 0.00 & 0.00   & 0.00 \\ 
\bottomrule
\end{tabular}

\label{tab:mitigation_1-3}
\end{table}

Table \ref{tab:mitigation_1-3} presents the comparison of CBS under different mitigation strategies across three datasets. 

For the Credit dataset, mitigation strategies lead to noticeable improvements. The CBS for Sex decreases from 0.21 (baseline) to 0.15 (FS) and 0.10 (CoT), and is completely eliminated under FS-CoT (0.00). 

In the Adult dataset, mitigation effects vary across attributes. For Sex, CBS is reduced substantially from 0.78 to 0.41 (FS) and 0.05 (CoT), and further to 0.00 under FS-CoT. Similarly, Marital Status shows a strong reduction from 0.93 to 0.79 (FS), 0.08 (CoT), and 0.02 (FS-CoT). However, for attributes such as Age, the reduction is less obvious under FS and CoT alone (0.99 to 0.92 and 0.89), while FS-CoT again achieves a significant decrease (0.06). For Race and Region, CBS remains at 0.00 across all settings, indicating no observable bias in these attributes.

For the COMPAS dataset, similar patterns are observed. The CBS for Sex decreases from 0.74 to 0.42 (FS) and 0.11 (CoT), and is eliminated under FS-CoT (0.00). For Race, bias is already low in the baseline (0.07) and is further reduced to 0.01 (FS) and 0.00 (CoT and FS-CoT). In contrast, Age does not benefit from simple mitigation strategies, with CBS remaining high, while FS-CoT significantly reduces it to 0.07.

\begin{tcolorbox}[
  colback=white,
  colframe=gray,
  boxrule=0.4pt,
  left=6pt,
  right=6pt,
  top=6pt,
  bottom=6pt
]
\textbf{Finding 8.} 
The mitigation results further support our Finding 7 that the treatment of protected attributes is context-dependent in LLMs. Attributes that are widely recognised as protected are consistently mitigated across strategies, while less explicitly recognised attributes require additional guidance, such as reasoning-based prompts, to achieve effective bias reduction. 
\end{tcolorbox}

\begin{tcolorbox}[
  colback=white,
  colframe=gray,
  boxrule=0.4pt,
  left=6pt,
  right=6pt,
  top=6pt,
  bottom=6pt
]
\textbf{Finding 9.} 
Providing explicit reasoning examples through FS-CoT significantly improves bias mitigation across all attributes, including those that are not commonly recognised as protected.
\end{tcolorbox}

\begin{figure}[!htpb]
	\centering

    \includegraphics[width=0.9\textwidth]{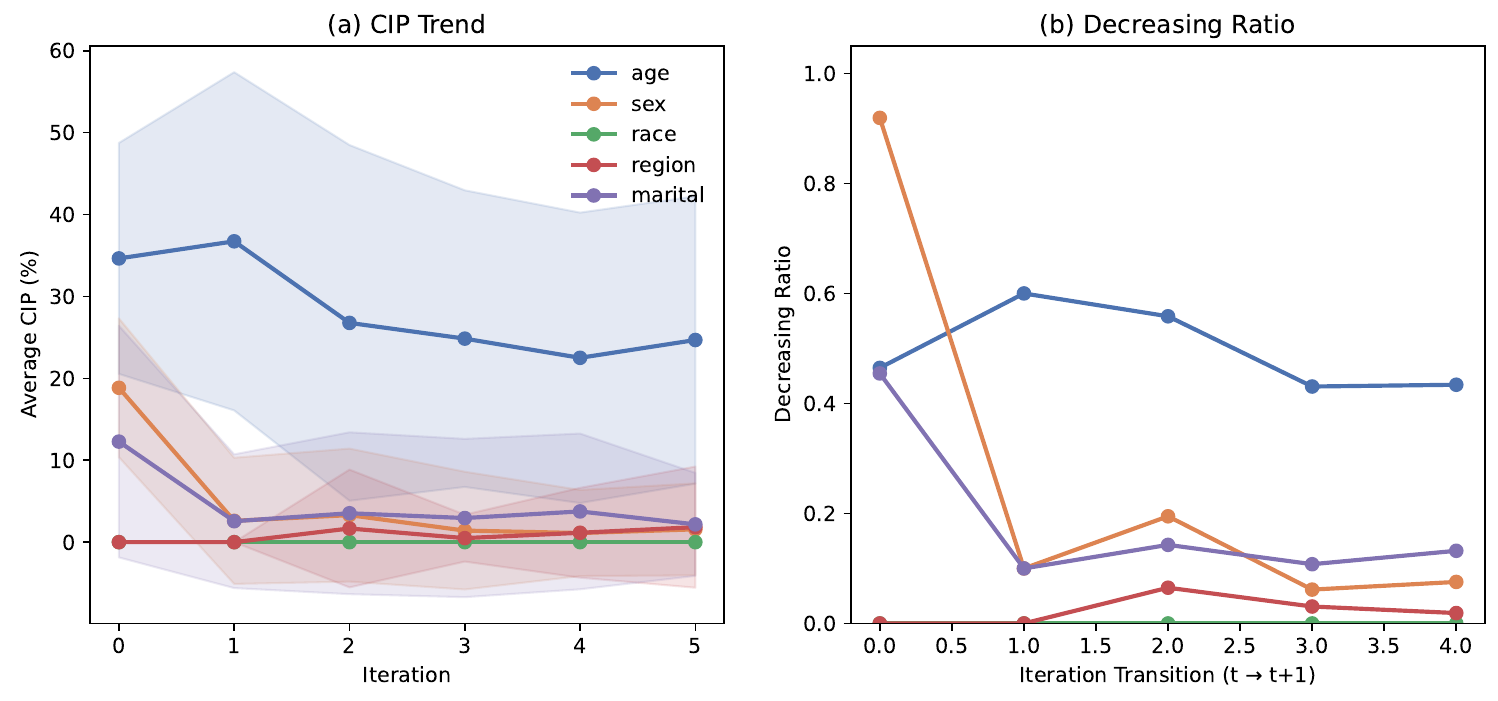}
  
	\caption{Iterative CIP Trends on the Adult Dataset}
   
    \label{fig:cip}
\end{figure}

As we mentioned before, we plan to investigate CIP value trends across iterations in MAS. Given that the Adult dataset contains multiple protected attributes, we use it as a representative case and present the results in Fig. \ref{fig:cip}.

Fig. \ref{fig:cip}(a) shows the average CIP for each protected attribute at each iteration. The CIP generally decreases, indicating that the proposed MAS effectively reduces bias over iterations. The most significant reductions occur in the early stages, particularly for Sex and Marital, while later iterations show more gradual changes. The shaded regions indicate the standard deviation across different functions, providing evidence of the consistency of the mitigation process. They show that although the average CIP decreases, the effectiveness of refinement varies across functions and attributes.

Fig. \ref{fig:cip}(b) presents the decreasing ratio between iterations, reflecting the extent of bias reduction at each refinement step. The decreasing ratio is highest in the initial iterations and decreases in later stages, confirming that most of the bias reduction occurs early in the process. 

\begin{tcolorbox}[
  colback=white,
  colframe=gray,
  boxrule=0.4pt,
  left=6pt,
  right=6pt,
  top=6pt,
  bottom=6pt
]
\textbf{Finding 10.} 
Although some fluctuations are observed, the overall trend indicates that bias is progressively reduced without substantial increases. This means the process remains generally stable toward improved fairness.
\end{tcolorbox}

\begin{tcolorbox}[
  colback=white,
  colframe=black,
  colback=black!5!white,
  boxrule=0.3pt,
  left=2pt,
  right=2pt,
  top=2pt,
  bottom=2pt
]
\noindent \textbf{Answer to RQ6:} 
Bias in LLM-generated decision functions can be mitigated, but effectiveness varies across methods and attributes. FS and CoT provide limited improvement, while FS-CoT achieves the most reliable improvement. MAS further reduce bias through iterative refinement, but their effectiveness is less stable and depends on the attribute.
\end{tcolorbox}
\subsection{Threats to internal validity}
Threats to internal validity mainly arise from the inherent variability of LLM-generated code, the definition of bias, and the iterative mitigation process. Given the same task description and prompt structure, an LLM may generate different decision functions due to its generative nature or imperfect task understanding. This variability is further amplified in the multi-agent mitigation setting, where iterative refinement may introduce additional stochasticity across iterations.
To mitigate the influence of such variability, we generate 100 functions per dataset and per prompt, then compute bias prevalence as proportions for all RQs, and report average values across functions for RQ3 and CIP trends for RQ6. This design reduces the impact of outlier functions. In addition, in both our bias measurement framework and the MAS, bias is detected using a rule-based, deterministic evaluation procedure based on the CIP, thereby reducing subjectivity and avoiding the randomness introduced by model training or sampling. 

\subsection{Threats to external validity}
Threats to external validity relate to the generalizability of our findings. Our experiments focus on a limited set of datasets, tasks, and LLMs, and bias behaviours may differ in other application domains or with future models. In particular, the attribute influence ranking reported in RQ3 is specific to the evaluated dataset and may not generalise to other data distributions or application contexts. Moreover, our analysis is based on deterministic decision functions generated under controlled prompts, whereas real-world LLM usage may involve more interactive or stochastic scenarios. Therefore, the results should be interpreted as evidence of the applicability of the proposed bias evaluation framework rather than exhaustive conclusions about all LLM-based systems.
\section{Related Work}\label{sec:relatedwork}
\noindent\textbf{Bias and fairness in Machine Learning.} With the development of machine learning and concerns about discrimination, bias detection in ML has gained widespread attention in the field~\cite{Li:2022}~\cite{Nabi:2019}. Early work by Friedman and Nissenbaum \cite{Friedman:1996} systematically categorised sources of bias in computer systems, laying the foundation. Subsequent studies proposed formal definitions of fairness~\cite{Lohaus:2020}~\cite{Mcnamara:2017}, such as individual fairness~\cite{Dwork:2012} and group fairness~\cite{Kusner:2017}. Mehrabi~\cite{Mehrabi:2021} surveyed common sources of bias in machine learning, while Barocas~\cite{Barocas:2023} and Verma et al.~\cite{Verma:2018} emphasised controlled evaluation and formal bias definitions as the foundation for fairness analysis.

\noindent\textbf{Bias in Large Language Models.} Some studies~\cite{Caliskan:2017}~\cite{Bolukbasi:2016}~\cite{Mary:2019}~\cite{Monteiro:2021} have shown that LLMs can inherit and amplify social biases present in their training data. Bender et al.~\cite{Bender:2021} emphasised concerns about bias, misuse, and a lack of accountability, particularly when protected personal attributes are involved. 

Subsequent work has focused on systematically evaluating their social biases. Some studies have proposed evaluation frameworks for language models to examine fairness and bias across multiple tasks~\cite{Liang:2022}~\cite{Nadeem:2021}. Similarly, the technical reports on GPT-3 and GPT-4~\cite{Achiam:2023} acknowledge bias in model outputs and emphasise the importance of systematic bias evaluation and mitigation. 
In parallel, fairness testing tools such as AIF360~\cite{Bellamy:2019} and Fairlearn~\cite{Bird:2020} have been developed to assess and mitigate bias in machine learning systems. While these tools~\cite{Lowy:2021} provide standardised fairness metrics and evaluation for bias in LLMs, they are primarily designed for natural language tasks rather than LLM-generated code.

\noindent\textbf{Bias in LLM-generated Code.} Liu \cite{Liu:2023} presented a new paradigm for constructing code hints, experiments on three pre-trained code generation models(Codex, InCoder, and CodeGen), and succeeded in revealing social bias in code generation models. However, beyond these three pre-training models, ChatGPT and Gemini are two of the most popular and widely used code-generation tools, so our research aims to evaluate bias behaviour on them. 
\section{Conclusion}\label{sec:conclusion}
This paper presents a systematic framework for evaluating bias in LLM-generated code. By varying protected attributes, we measure both bias prevalence and attribute-level influence, and show that the framework generalises across different LLMs. Our results further show that while generation settings, including prompt design and the integration of external knowledge (e.g., web search), can influence bias, they are insufficient to consistently and reliably mitigate it. To address this, we explore several lightweight mitigation strategies. Approaches incorporating reasoning or iterative feedback achieve more consistent bias reduction, although bias is not completely eliminated. Future work will extend this framework by jointly evaluating utility and fairness, and by exploring iterative refinement strategies for improving decision quality.
\section{Data Availability} \label{sec:data_availability}

The dataset and code can be found at:~\cite{BiasMeasurementWebsite}

\bibliography{sec/reference}

\end{document}